\documentclass[useAMS,usenatbib]{mn2e}
\usepackage{epsfig}
\usepackage{times}
\usepackage[fleqn]{amsmath}
\usepackage{amssymb}
\usepackage{amsbsy}

\title[Non-parametric foreground subtraction]{Non-parametric
  foreground subtraction for 21cm epoch of reionization experiments}

\author[G. J. A. Harker et al.]{Geraint Harker,$^{1}$\thanks{E-mail:
  harker@astro.rug.nl} Saleem Zaroubi,$^{1}$ Gianni Bernardi,$^{1}$
  Michiel A. Brentjens,$^{2}$ \newauthor A. G. de Bruyn,$^{1,2}$
  Benedetta Ciardi,$^{3}$ Vibor Jeli\'c,$^{1}$ Leon
  V. E. Koopmans,$^{1}$ \newauthor Panagiotis Labropoulos,$^{1}$
  Garrelt Mellema,$^{4}$ Andr\'e Offringa,$^{1}$ V. N. Pandey,$^{1}$
  \newauthor Joop Schaye,$^{5}$ Rajat M. Thomas$^{1}$ and Sarod
  Yatawatta$^{1}$\\
  $^{1}$Kapteyn Astronomical Institute, University of Groningen, PO
  Box 800, 9700AV Groningen, the Netherlands\\
  $^{2}$ASTRON, Postbus 2, 7990AA Dwingeloo, the Netherlands\\
  $^{3}$Max-Planck Institute for Astrophysics,
  Karl-Schwarzschild-Stra\ss e 1, 85748 Garching, Germany\\
  $^{4}$Department of Astronomy and Oskar Klein Centre for
  Cosmoparticle Physics, AlbaNova, Stockholm University, SE-106 91
  Stockholm, Sweden\\
  $^{5}$Leiden Observatory, Leiden University, PO Box 9513, 2300RA
  Leiden, the Netherlands}

\begin{document}

\date{\today}

\maketitle

\begin{abstract}
One of the problems facing experiments designed to detect redshifted
21cm emission from the epoch of reionization (EoR) is the presence of
foregrounds which exceed the cosmological signal in intensity by
orders of magnitude. When fitting them so that they can be removed, we
must be careful to minimize `overfitting', in which we fit away some
of the cosmological signal, and `underfitting', in which real features
of the foregrounds cannot be captured by the fit, polluting the signal
reconstruction. We argue that in principle it would be better to fit
the foregrounds non-parametrically -- allowing the data to determine
their shape -- rather than selecting some functional form in advance
and then fitting its parameters. Non-parametric fits often suffer from
other problems, however. We discuss these before suggesting a
non-parametric method, Wp smoothing, which seems to avoid some of
them.

After outlining the principles of Wp smoothing we describe an
algorithm used to implement it. Some useful results for implementing
an alternative algorithm are given in an appendix. We apply Wp
smoothing to a synthetic data cube for the Low Frequency Array (LOFAR)
EoR experiment. This cube includes realistic models for the signal,
foregrounds, instrumental response and noise. The performance of Wp
smoothing, measured by the extent to which it is able to recover the
variance of the cosmological signal and to which it avoids the fitting
residuals being polluted by leakage of power from the foregrounds, is
compared to that of a parametric fit, and to another non-parametric
method (smoothing splines). We find that Wp smoothing is superior to
smoothing splines for our application, and is competitive with
parametric methods even though in the latter case we may choose the
functional form of the fit with advance knowledge of the simulated
foregrounds. Finally, we discuss how the quality of the fit is
affected by the frequency resolution and range, by the characteristics
of the cosmological signal and by edge effects.
\end{abstract}


\section{Introduction}\label{sec:intro}

Several current and upcoming facilities (e.g.\ GMRT,\footnotemark\
MWA,\footnotemark\ LOFAR,\footnotemark\ 21CMA,\footnotemark\
PAPER,\footnotemark\ SKA\footnotemark) aim to detect redshifted 21cm
line emission from the epoch of reionization (EoR). One problem all
such experiments face is to disentangle the desired cosmological
signal (CS) from foregrounds which are orders of magnitude larger
\citep{SHA99}. It is hoped and expected that these foregrounds will be
smooth as a function of frequency, while the signal we wish to detect will
fluctuate on small scales \footnotetext[1]{Giant Metrewave Telescope,
http://www.gmrt.ncra.tifr.res.in/}\footnotetext[2]{Murchison Widefield
Array,
http://www.haystack.mit.edu/ast/arrays/mwa/}\footnotetext[3]{Low
Frequency Array, http://www.lofar.org/}\footnotetext[4]{21 Centimeter
Array, http://web.phys.cmu.edu/\~{}past/}\footnotetext[5]{Precision
Array to Probe the EoR,
http://astro.berkeley.edu/\~{}dbacker/eor/}\footnotetext[6]{Square
Kilometre Array, http://www.skatelescope.org/}(\citealt{SHA99};
\citealt{DIM02}; \citealt{OH03}; \citealt*{ZAL04}). If we subtract the
smooth component, then the residual will contain contributions from
fitting errors (hopefully small), the signal (hopefully largely
intact) and noise. Because 21cm emission is line emission, redshift
information translates into spatial information along the line of
sight (modulo redshift space distortions), thus in principle allowing
us to carry out 21cm tomography. In practice, however, for the current
generation of instruments such as LOFAR or MWA, the noise per
resolution element is expected to exceed the signal by a factor of
several, and a spatial resolution element is expected to be of order
the size of interesting features of the signal. Current experiments
therefore aim to measure statistics such as the global signature of
reionization or the power spectrum of 21cm emission. We wish to find
foreground subtraction algorithms which do not introduce a large bias
into these statistics or make the properties of the noise more
awkward.

In this paper we propose a non-parametric technique, `Wp smoothing'
\citep{MAC93,MAC95}, as a way to fit the foregrounds. This method
involves calculating a least-squares fit to the brightness temperature
as a function of frequency along each line of sight, subject to a
penalty on changes in curvature.

Our approach differs from the one usually taken in the recent
literature on foreground subtraction, in which, at some point, a
specific functional form for the foregrounds is assumed. For example,
\citet*{SAN05} assumed that the foregrounds consisted of four
components with spectra which were power laws with spatially constant
spectral index, and fit the parameters of these power laws, along with
parameters desribing how the foregrounds correlate between different
frequencies, simultaneously with estimating the power spectrum of the
CS. Progressing to a model in which the spectral index changes with
position, \citet{XWANG06} fit the log intensity along each line of
sight with a second-order polynomial in $\log\nu$.  Polynomial fitting
in frequency or log frequency is rather popular (\citealt*{MOR06};
\citealt{MCQ06}; \citealt*{GLE08}; \citealt*{BOW09}), and we have used
it in previous work in which we tested our ability to extract
properties of the EoR signal from a simulated data cube with realistic
foregrounds \citep{JEL08}. We also use it in
Section~\ref{subsec:comparison} in order to compare with our
non-parametric approach. This permits reasonable recovery of the CS,
but leaves us with some concerns.  Firstly, the function has to be
carefully chosen, both to be able to capture the shape of the
foregrounds and to have the right amount of freedom: for example, in
our simulations a second-order polynomial has insufficient freedom and
produces biased fits (the difference from the results of
\citealt{XWANG06} may be due to the larger frequency range studied
here), while a fourth-order polynomial has too much freedom and `fits
out' some of the signal.  Secondly, we knew the original foregrounds
by construction and could use this fact to test our recovery, whereas
in the real observations the data themselves will provide the best
estimate of the foregrounds.  This latter point suggests using a
non-parametric fit in which the shape of the fit is `chosen' by the
data.

We must also be careful in our selection of a non-parametric method,
however.  One could consider using, for example, `smoothing splines':
piecewise polynomial functions which minimize the sum of the squared
residuals and a term which measures the integrated squared curvature.
A smoothing parameter, $p$, adjusts the relative weight of the
least-squares term and the curvature term.  If the least-squares term
is given a large weight then the smoothing spline becomes an
interpolating function, passing through all the data points, which is
clearly undesirable.  As the curvature term is given larger weight,
the smoothing spline becomes closer to being a straight line, which
leads to a systematic bias in the estimate of the foregrounds if they
have any curvature.  In practice, this would not be a problem if there
were some intermediate value of the smoothing parameter which led to
acceptable fits, or if there were a well defined procedure for
choosing a smoothing parameter for a given problem.  We have found,
though, that there is no value for which we do not see overfitting,
large bias or both. The limiting behaviour of Wp smoothing as its
smoothing parameter $\lambda$ takes very small or very large values is
much more suited to our foreground fitting problem, and we will show
that it produces good fits for a wide range of values of $\lambda$.

The spectral fitting which is the primary focus of this paper
constitutes only one step in the foreground subtraction, which in
general is a procedure with several stages which may interact
\citep{MOR04,MOR06}. We assume that the first stage, the subtraction
of bright point sources, has already been carried out on our data
cubes. In the stage following the spectral fitting, the different
symmetries possessed by the fitting errors and the cosmological signal
may be exploited when performing parameter estimation. We do not dwell
on that here, though we touch briefly on the use of the
non-Gaussianity of the CS \citep[e.g][]{FUR04b} to enhance the
recovery of a signal when we look at the skewness of the residual maps
in Section~\ref{subsec:chooselambda}.

A comparison of results on the quality of foreground subtraction
using several methods, including Wp smoothing, polynomial fitting and
smoothing splines, may be found in Section~\ref{sec:res}.  Here we
show that Wp smoothing overcomes the problems posed by parametric fits
and by other non-parametric methods. As we have just noted, for
Wp smoothing we must specify the value of a smoothing parameter,
$\lambda$.  We suggest a way of choosing $\lambda$ and examine its
effects on our results, then show that some statistical properties of
the signal can be extracted well after removal of the foregrounds
using Wp smoothing.  Before that, in Section~\ref{sec:sims}, we start
by briefly describing the simulations on which our results are
based. Then, in Section~\ref{sec:method}, we lay some groundwork by
sketching the mathematical basis of our method, and showing how we
solve the differential equation which the Wp smoother fulfils. An
appendix gives some intermediate results that may be useful for others
who may wish to solve the equation by a different route. Some
conclusions are offered in Section~\ref{sec:conc}.

\section{Simulations of EoR data}\label{sec:sims}

We test our fitting techniques on the same synthetic data cubes as
we have used in previous work \citep{SKE_09}. They have three
components: the CS, the foregrounds and the noise. The data cube
consists of spatial slices of $256^2$ pixels, representing an
observing window with an angular size on the sky of $5\degr\times
5\degr$. This corresponds to a square of side $624\ h^{-1}\
\mathrm{Mpc}$ (comoving) at $z=10$ in the cosmology assumed in the
simulation. There are 170 such slices, spaced at intervals of $0.5\
\mathrm{MHz}$ in observing frequency between $115$ and $200\
\mathrm{MHz}$. These frequencies correspond to redshifts of the 21cm
line of between $11.35$ and $6.12$. At $150\ \mathrm{MHz}$, $\Delta
\nu=0.5\ \mathrm{MHz}$ corresponds to $\Delta z\approx 0.03$, or a
comoving radial distance of around $7\ h^{-1}\ \mathrm{Mpc}$.

We estimate the CS primarily using the simulation f250C of
\citet{ILI08}. The distribution of dark matter in a $100\ h^{-1}\
\mathrm{Mpc}$ box was followed using $1624^3$ particles on a $3248^3$
mesh, and the ionization fraction was then calculated in
post-processing on a $203^3$ mesh. The parameters of the assumed
cosmology were \hbox{($\Omega_\mathrm{m}$, $\Omega_\Lambda$,
$\Omega_\mathrm{b}$, $h$, $\sigma_8$, $n$)}$=$\hbox{(0.24, 0.76,
0.042, 0.73, 0.74, 0.95)}. A datacube of $203\times 203\times 3248$
points was generated from the periodic simulation boxes according to
the method described by \citet{MEL06b}, where the long dimension is
the frequency dimension. We then tiled copies of this cube in the
plane of the sky in order to fill our observing window, before
interpolating onto our $256\times 256\times 170$ grid.  The tiling
means that there are periodic repetitions in the CS in the plane of
the sky, which may introduce problems if we were to study spatial
statistics, for example the power spectrum. We do not study such
statistics here, however. We are interested mainly in how well the
signal is recovered given the foregrounds and noise, the maps of which
are generated for the full observing window and therefore have no
periodic repetition. Two pixels which receive their CS contribution
from the same pixel of the original CS map may none the less have very
different contributions from the foregrounds and noise.

For comparison with our results using f250C, in
Section~\ref{subsec:altsims} we study two simulations described by
\citet{THO09}. These use a one-dimensional radiative transfer code
\citep{THO08} in conjunction with a dark matter simulation of $512^3$
particles in a $100\ h^{-1}\ \mathrm{Mpc}$ box. They differ only in
the source properties: in one simulation it is assumed that the
Universe is reionized by QSOs, and in the other by stars. We label
these simulations `T-QSO' and `T-star' respectively. Data cubes are
derived from the periodic simulation boxes in a similar fashion as was
done for the f250C simulation.

We use the foreground simulations of \citet{JEL08}. The dominant
component in these simulations is the contribution from Galactic
diffuse synchrotron emission. This is calculated by first generating a
four-dimensional realization (three spatial dimensions and one
frequency dimension; see e.g. \citealt{SUN08} for recent constraints
on such realizations) and then integrating along the spatial direction
parallel to the line of sight to produce a three-dimensional datacube
(two spatial dimensions and one frequency dimension). Galactic
free-free emission is generated in a similar manner. Free-free
emission is expected to be much weaker than synchrotron at these
frequencies, contributing around one per cent of the total foreground
emission. None the less, on its own it would still dominate the 21cm
signal. The final Galactic contribution to the foreground maps comes
from supernova remnants: two such remnants, modelled as discs of
uniform surface brightness with a flux density, angular size and
spectral index drawn from an observational catalogue, are placed at
random positions on the map.  The extragalactic foregrounds comprise
two types of source: radio galaxies and radio clusters.  The radio
galaxies consist of star-forming galaxies as well as FR~I and FR~II
sources, with realistic flux density distributions and angular
clustering. Though they are modelled as uniform discs, they are in any
case almost point-like at the resolution of the LOFAR EoR
experiment. The radio clusters are also modelled as uniform discs,
with steep power-law spectra, and with sizes and positions taken from
an $N$-body simulation.

The main limitation of the foreground simulations used here is perhaps
the fact that we do not include Jeli{\'c} et al.'s modelling of the
polarization. Observational constraints on the levels of polarization
at the scales and frequencies of relevance to EoR experiments have
only recently been obtained \citep{PEN08,BER09}. Even in the absence
of a polarized signal from the sky, though, the intrinsically
polarized response of the antennas used in an experiment such as LOFAR
will compel us to consider polarization in a final analysis. Poor
polarization calibration may, for example, allow fluctuations in
polarization to leak into the unpolarized power, contaminating the EoR
signal. Such issues are not the focus of this paper, however, and so
our simplified model of the instrumental response also neglects
polarization.

We include the effects of the instrumental response of LOFAR on the
signal and foregrounds by performing a two-dimensional Fourier
transform on each image, multiplying by a sampling function which
describes how densely the interferometer baselines sample Fourier
space, and then performing an inverse transform. The density of
visibilities is calculated using the planned locations of the LOFAR
core stations which will be used for EoR studies, and assuming that
our observational window is located at declination $\delta=90^\circ$
and observed for four hours each night. In reality the LOFAR project
will observe several windows at different declinations, and may
integrate for longer each night: the observing plan is not yet
finalized. It should be borne in mind that the results of this paper
are confined to a single observing window, but that having several in
the final data set will allow important cross-checks as well as
improving the statistics. At present we use the same sampling
function at all frequencies; in reality, the `{\it uv} coverage' (the
region of the Fourier plane where the sampling function is not zero)
changes with frequency, so this amounts to ignoring information from
parts of the Fourier plane which are not sampled at all
frequencies. If we adopt a strict criterion whereby we discard
{\it uv} points which have data missing at any frequency, then with
the {\it uv} plane gridding and frequency coverage adopted in this
paper less than 20 per cent of the data would have to be discarded. If
the frequency range is shortened to $115$--$180\ \mathrm{MHz}$, and if
we relax our criterion so that {\it uv} points at which there are
measurements in at least 95 per cent of the frequency channels are
included, then the proportion of data which must be discarded goes
down to less than ten per cent. Clearly it would be desirable to
reduce this still further: for example, \citet{LIU09b} describe a
method which is claimed to alleviate some of the problems associated
with changing {\it uv} coverage, though it is not yet clear how their
method generalizes to nonlinear fitting techniques such as those
studied in this paper.

Noise images are produced by generating uncorrelated Gaussian noise at
grid points in the Fourier plane where the sampling function is not
zero, transforming to the image plane, and then normalizing this noise
image so that it has the correct {\it rms}. The noise {\it rms} is
calculated as in \citet{JEL08} assuming 400 hours of
integration, and includes a frequency-dependent part from the sky
(scaling roughly as $\nu^{-2.55}$) and a frequency-independent
part from the receivers, such that at $150\ \mathrm{MHz}$ it has a
value of $52\ \mathrm{mK}$. The frequency dependence of the noise
results from the frequency dependence of the system temperature, which
is modelled as $T_\mathrm{sys}=140+60(\nu/300\ \mathrm{MHz})^{-2.55}\
\mathrm{K}$.

The instrumental corruptions introduced by the observing process will
clearly be rather more complex than we have assumed here;
\citet{LAB09} discuss this in more detail, with a view towards
developing a complete end-to-end model of the effects on the signal of
foregrounds, the atmosphere and the instrument. The weighting of
points in the {\it uv} plane used to generate our image cubes may also
affect the foreground subtraction \citep{BOW09}, but we have not
investigated the effect of such changes here.

\section{The Wp method}\label{sec:method}

In this section we provide some justification for trying Wp smoothing
as a foreground fitting technique and briefly review the relevant
mathematical results given by \citet{MAC93,MAC95}. We then describe
our algorithm for implementing Wp smoothing.

\subsection{Background}\label{subsec:bg}

If pressed to explain what one meant by trying to find a `smooth'
curve that fit some data set, one might be tempted to say, for
example, that the curve had no `wiggles'.  A function with constant
curvature might well be considered extremely smooth in this sense. In
the case of the smoothing splines mentioned in the introduction,
however, the roughness of a curve is given by its integrated squared
curvature.  By this measure, a function with constant moderate
curvature could well be computed as being less smooth than an almost
straight line with small wiggles superimposed. This is the motivation
for considering, instead, the integrated \emph{change} of curvature.

To be more precise, suppose we have a set of observations
$\{(x_1,y_1),(x_2,y_2),\ldots,(x_n,y_n)\}$ which we wish to fit with a
smooth function $f(x)$. Each $y_i$ may have an associated error,
$\sigma_i$. In our context, the $x_i$ are a series of observing
frequencies and the $y_i$ the corresponding differential brightness
temperature, all at a given point on the sky. $\sigma_i$ is the {\it
rms} noise in the map at frequency $x_i$. Given a function $f(x)$, its
curvature, defined as the reciprocal of the radius of curvature, is
given by $\kappa(x)=f''(x)(1+f'(x)^2)^{-3/2}$, and its standardized
change of curvature (or change of log curvature) is given by
\begin{equation}
\frac{\kappa '}{\kappa}=\frac{f'''}{f''}-3\frac{f'f''}{1+f'^2}\approx
\frac{f'''}{f''}\ .\label{eqn:curvdef}
\end{equation}
The approximation shown holds exactly at local extrema ($f'=0$) and at
inflection points ($f''=0$) and is adopted for convenience.

The first thing to note is that the standardized change of curvature
becomes singular at inflection points. Thus the number of inflection
points that a function possesses is the most important determinant of
its roughness, and we need some sort of procedure to specify the
number and position of the inflection points of our smoothing
function. Once this is done, we need a way to measure the roughness
`apart from inflection points' to finally specify the function. The
importance of inflection points is reflected in the name of the
method, `Wp' being short for the German word `Wendepunkt', meaning
`inflection point'.

Suppose, then, that the inflection points $w_j$, $j=1,2,\ldots,n_w$,
are given. Then we may write
\begin{equation}
f''(x)=p_\mathbf{w}(x)\mathrm{e}^{h_f(x)} ,\label{eqn:f2deriv}
\end{equation}
where
\begin{equation}
p_\mathbf{w}(x)\equiv s_f(x-w_1)(x-w_2)\ldots(x-w_{n_w})\ ,
\end{equation}
$s_f=\pm 1$ and $h_f$ is a function as many times differentiable as
$f''$. Now,
\begin{equation}
\frac{f'''}{f''}=\frac{\mathrm{d}}{\mathrm{d}x}\log f''=(\log
p_\mathbf{w})'+h_f'
\end{equation}
or, rearranging,
\begin{equation}
h_f'=\frac{f'''}{f''}-\sum_{j=1}^{n_w}\frac{1}{x-w_j}\ .
\end{equation}
This separates our measure of roughness into a part which depends on
the number and position of the inflection points and a part which
depends on the other properties of $f$.

We may then express the smoothing problem, given the position of the
inflection points, as follows.  We wish to find the function $f$ which
minimizes
\begin{equation}
\sum_{i=1}^n \rho_i(y_i-f(x_i)) +
\lambda\int_{x_1}^{x_n}h_f'(t)^2\mathrm{d}t\ ,
\label{eqn:minimizes}
\end{equation}
where $\lambda$ is a Lagrange multiplier, the integral term measures
the change in curvature `apart from inflection points' and the
function $\rho_i$ determines the size of the penalty incurred when
$f(x_i)$ deviates from $y_i$. For simple least-squares minimization,
for example, $\rho_i(\delta)=\frac{1}{2}\delta^2$ for all $i$, where
$\delta$ is the difference between the data and the fitting function.

The solution of this minimization problem must then satisfy an
ordinary differential equation (ODE) and boundary conditions derived
by \citet{MAC93,MAC95}, who also considered more general cases, for
example using higher derivatives of $h_f$ in the integral term. The
ODE found by M\"achler for the Wp smoothing case is as follows:
\begin{equation}
h_f''=p_\mathbf{w}\mathrm{e}^{h_f}L_f\ ,
\label{eqn:mainode}
\end{equation}
where, using the notation $a_{\scriptscriptstyle +}=\mathrm{max}(0,a)$,
\begin{equation}
L_f(x)=-\frac{1}{2\lambda}\sum_{i=1}^{n}(x-x_i)_{\scriptscriptstyle +}\psi_i[y_i-f(x_i)]
\end{equation}
and
$\psi_i(\delta)=\frac{\mathrm{d}}{\mathrm{d}\delta}\rho_i(\delta)$.
The solution must satisfy some simple boundary conditions,
\begin{equation}
h_f'(x_1)=h_f'(x_n)=0\ ,
\label{eqn:bcdef1}
\end{equation}
as well as some rather more problematic boundary conditions,
\begin{equation}
\sum_i\psi_i(y_i-f(x_i))=\sum_ix_i\psi_i[y_i-f(x_i)]=0\ .
\label{eqn:bcdef2}
\end{equation}
We may write $\psi_i$ explicitly as $\psi_i(\delta)=\delta$ for least
squares, or, taking the errors into account, as
$\psi_i(\delta)=\delta/\sigma_i$. Equivalently, each data point is
associated with a weight $c_i=1/\sigma_i$; this is our default
weighting scheme, but we consider other choices in
Section~\ref{subsec:weights}. Alternatively, a more robust method may use
\begin{equation}
\psi_i(\delta)=\begin{cases}
C & \text{if $\delta/\sigma_i>C$ ,} \\
\delta/\sigma_i & \text{if $|\delta/\sigma_i|\leq C$ ,} \\
-C & \text{if $\delta/\sigma_i<-C$}
\end{cases}
\end{equation}
for some $C>0$. When working with our simulations this
complication is unnecessary, since the synthetic data cubes have no
outliers by construction. We mention it here only for completeness,
since a robust method may become necessary for dealing with more
realistic simulations and with the real data, and to illustrate the
point that the method can deal with a general choice of penalty
function.

Not only are the boundary conditions problematic, but the ODE itself,
equation~\eqref{eqn:mainode}, includes on the right-hand side a
contribution from $f(x_i)$ for each $x_i$, meaning that the equation
is not in the `standard form' assumed by off-the-shelf solvers for
boundary value problems (BVPs).

Recall that the minimization is performed with $s_f$ and $\{w_i\}$
fixed. To apply the procedure to an arbitrary data set, then, requires
a further minimization over the number and position of the inflection
points. We therefore require some method to give a starting
approximation for $f$, $f'$, $h_f$, $h_f'$, $n_w$, $\{w_i\}$ and
$s_f$. For our particular application we need not consider arbitrary
data sets: the properties of the foregrounds seem to allow us to
achieve acceptable fits with $n_w=0$. This might be expected if the
foregrounds were to consist of a superposition of power laws with
varying spectral index, for example arising from different sources
along the line of sight. We therefore impose this condition throughout
this paper (with one small exception discussed in the following
subsection) and do not examine how to perform a minimization over
$n_w$ and $\{w_i\}$.

In principle we should also like some method to choose the Lagrange
multiplier, $\lambda$. Wp smoothing remains well defined for
$\lambda\to 0$ and $\lambda\to\infty$. Indeed, an attractive feature
of the method is that for $\lambda\to 0$, $f$ does not become an
interpolating function as happened for smoothing splines: rather, it
becomes the best-fitting function having the given inflection
points. Meanwhile, for $\lambda\to\infty$, $f$ becomes the
best-fitting polynomial of degree $n_w+2$ with the given inflection
points, rather than becoming a straight line which automatically
underestimates the curvature.

\citet{MAC93} suggests using the autocorrelation function of the
residuals to estimate $\lambda$, reducing $\lambda$ from a large value
in stages until the residuals become uncorrelated. This could be
problematic for our application, since there may be real correlations
in the noise between frequency bands due to the CS we
aim to find. 

We might note instead that because $\lambda$ controls the degree of
regularization we apply during the fitting, with smaller $\lambda$
affording a greater degree of freedom in the functional form, a choice
of $\lambda$ expresses our prior knowledge of how smooth we expect the
foregrounds to be. If that knowledge is uncertain, a fully consistent
approach would be to estimate what level of freedom is justified by
the data themselves. Such a framework could also encompass the choice
of $n_w$, which may be viewed as a more important regularization
parameter. This sort of problem, and the topic of mixed signal
separation in general, is of course the subject of an extensive
literature in information theory and Bayesian
inference. Unfortunately, Wp smoothing seems to present a rather
awkward case for such methods. Since it is already quite
computationally expensive to calculate the Wp smoothing solution for
even a single value of $\lambda$, we have not chosen to go via this
route.

We have instead taken a more heuristic approach, smoothing using
different values for $\lambda$ and using our knowledge of the
simulated foregrounds to test the quality of the fit according to
various criteria. We detail these criteria, and use our results on
simulated datacubes to choose the value of $\lambda$ used for the
subsequent parts of the paper, in Section~\ref{subsec:chooselambda}.

In the following subsection we give some details of the algorithm we
use to solve equation~\eqref{eqn:mainode}.  A reader uninterested in
these details should skip directly to Section~\ref{sec:res}, in which
we describe our results.

\subsection{Algorithm}\label{subsec:alg}

An algorithm to solve equation~\eqref{eqn:mainode} subject to the
boundary conditions given by equations~\eqref{eqn:bcdef1}
and~\eqref{eqn:bcdef2} is reportedly given by \citet{MAC89}. Since there
is no publicly available implementation of this algorithm, and since
we will not deal with the most general case, we have experimented with
different approaches. The first is to rewrite the differential
equation as in Appendix~\ref{sec:appa}, such that it can be solved by
a standard BVP solver. The second, which we have found to be faster
and more stable (though giving identical results) is to discretize
the differential equation into a finite difference equation defined on
a grid, and then solve the resulting algebraic system using standard
methods.

We choose a mesh such that the abscissae of the data points are also
mesh points. That is, we have a mesh $X_1,X_2,\ldots,X_N$, where $N\ge
n$, and where $X_{m_i}=x_i$ for $i=1,\ldots n$, with $m_1=1$ and
$m_n=N$. A mesh with two additional points between each pair of data
points (that is, with $N=3n-2$) seems to be adequate, in that adding
more mesh points does not change the solution at the position of the
data points to high accuracy.

Let $f(X_i)=f_i$ and $h(X_i)=h_i$ (which implies that
\hbox{$f(x_i)=f_{m_i}$}). Further, let
$\Delta_j=(X_{j+1}-X_j)(X_j-X_{j-1})$. Then we may discretize
equation~\eqref{eqn:f2deriv} as
\begin{equation}
f_{j+1}-2f_j+f_{j-1}-\Delta_jp_\mathbf{w}(X_j)\mathrm{e}^{h_j}=0\ . \label{eqn:ffd}
\end{equation}
Similarly, we may rewrite equation~\eqref{eqn:mainode} as
\begin{equation}
\begin{split}
0={}&h_{j+1}-2h_j+h_{j-1} \\
&-\Delta_jp_\mathbf{w}(X_j)\mathrm{e}^{h_j}\left[
-\frac{1}{2\lambda}\sum_{i=1}^n(X_j-x_i)_{\scriptscriptstyle
  +}\psi_i(y_i-f_{m_i})\right]\ ,
\label{eqn:hfd}
\end{split}
\end{equation}
where in each case the index $j$ runs from 2 to $N-1$. The boundary
conditions of equation~\eqref{eqn:bcdef1} become
\begin{equation}
h_2-h_1 = h_N-h_{N-1} = 0 \ ,\label{fdbc1}
\end{equation}
while those of equation~\eqref{eqn:bcdef2} become
\begin{equation}
\sum_i\psi_i(y_i-f_{m_i})=\sum_ix_i\psi_i(y_i-f_{m_i})=0 \
.\label{eqn:fdbc2}
\end{equation}
We solve the system of equations~\eqref{eqn:ffd}--\eqref{eqn:fdbc2} using
the \textsc{matlab} routine `fsolve'. Our method is therefore
essentially a relaxation scheme, but one in which the unusual form of
equations~\eqref{eqn:hfd} and~\eqref{eqn:fdbc2} does not allow us to take
the shortcuts used by standard relaxation schemes, which exploit the
special form of algebraic systems arising from finite difference
schemes.

The initial guess for the solution is also important, and a poor guess
can greatly increase the execution time. A method for finding an
initial guess for a generic dataset would need to provide an estimate
of the number and position of the inflection points. We have found,
however, that we can fit the foregrounds using estimates with no
inflection points -- or, to put it another way, no wiggles --
i.e. $n_w=0$. Imposing this condition simplifies the problem.  It is
convenient to provide an initial guess for $f$ which also has no
inflection points within the range being fitted, and we have found
that using a power law works reasonably well.

In Fig.~\ref{fig:onelos} we have shown a fit for one line of
sight using $n_w=1$. To produce this fit we performed a further
minimization of the penalty function $\sum_{i=1}^n\rho_i(y_i-f(x_i))$
over the position of the inflection point, $w_1$. Lacking a method to
give an initial estimate for $w_1$, we used a simple golden section
search algorithm with $w_1$ constrained only to lie somewhere in the
frequency range spanned by the data. For all values of $w_1$ we used
the same initial guess for $f$ as for the $n_w=0$ case, i.e. a power
law. We have not attempted to provide an initial guess with an
inflection point in the right place, but fortunately the execution
time is not too severely affected. The extra minimization step
requires us to solve our algebraic system for more than 50 values of
$w_1$ in this instance.

Using this scheme, fitting the foregrounds using $n_w=0$ for one line
of sight for our fiducial value of $\lambda$ (see
Section~\ref{subsec:chooselambda}) usually takes less than one second
on a typical workstation, though some awkward lines of sight may
take tens of seconds. Going to smaller $\lambda$ does increase the
execution time, however. Our simulated data cube has $256^2$ lines of
sight, and since the fitting for each one is independent the
calculation can be trivially split between several processors, meaning
processing the cube typically takes of order a few hours on our setup.
Fitting using $n_w=1$ and minimizing over the position of the
inflection point increases the required time by approximately two
orders of magnitude, so we have not systematically studied the $n_w>0$
case.  We comment on this case when we show an example of an $n_w=1$
fit in Fig.~\ref{fig:onelos}, however.

\section{Results}\label{sec:res}

\begin{figure}
  \begin{center}
    \leavevmode \psfig{file=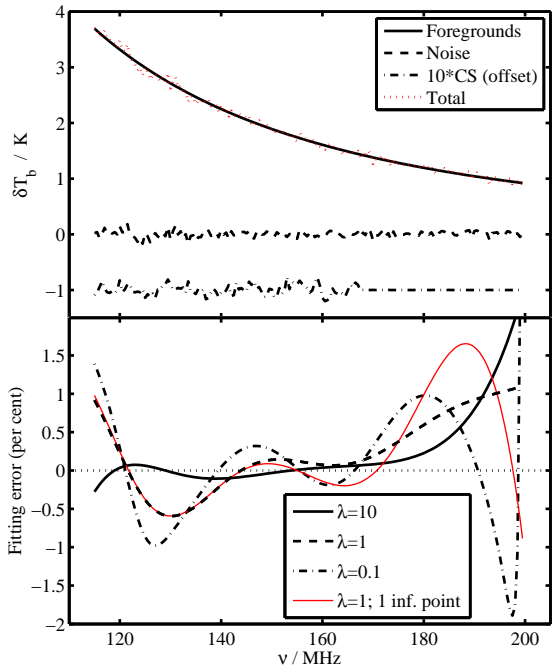,width=8cm}
    \caption{We show, in the top panel, the differential brightness
    temperature as a function of frequency for a particular line of
    sight. We also show the individual contributions to this total
    from the foregrounds, noise and CS. We have multiplied the size of
    the CS by a factor of 10, and offset the corresponding line by
    $-1\ \mathrm{K}$ for clarity. The bottom panel shows the
    difference between the Wp smoothing estimate of the foregrounds
    and their true value, expressed as a percentage, for three values
    of the smoothing parameter $\lambda$ with $n_w=0$, and for
    $\lambda=1$ with $n_w=1$. The inflection point in the latter case
    is at $w_1=188.4\ \mathrm{MHz}$.}\label{fig:onelos}
  \end{center}
\end{figure}

To illustrate the problem we are attacking, in the top panel of
Fig.~\ref{fig:onelos} we show the contribution to the differential
brightness temperature $\delta T_\mathrm{b}$ along an example line of
sight from the foregrounds, noise and CS.
For this particular line of sight, the total intensity is positive at
all frequencies. Because at each frequency the mean over all lines of
sight in one of our images must be zero, however (since an
interferometer cannot measure the mean, which for the foregrounds
could be as much as tens or hundreds of kelvin at these frequencies),
if we had chosen a different line of sight then we could have seen
$\delta T_\mathrm{b}<0$ for all $\nu$, or have seen some positive and
some negative values due to noise. The latter situation is atypical,
however, since the fluctuations in the foregrounds are of order a few
kelvin (making the line of sight shown in Fig.~\ref{fig:onelos} fairly
typical) whereas the noise fluctuations are of order tens of
millikelvin. The size of the CS has been increased by a factor of 10
for the plot, so that the fluctuations are visible; we have also
offset the line by $-1\ \mathrm{K}$ for clarity. The CS is very nearly
zero for $\nu\gtrsim 170\ \mathrm{MHz}$, owing to reionization.

The bottom panel of Fig.~\ref{fig:onelos} shows how well we estimate
the foregrounds by applying Wp smoothing to the total signal along
this line of sight, for three different values of $\lambda$ with
$n_w=0$, and for $\lambda=1$ with $n_w=1$. In the $n_w=1$ case, the
position of the inflection point is $w_1=188.4\ \mathrm{MHz}$. Though
no conclusions can be derived from a single line of sight, we can see
that accuracies of around one per cent or better are reached, and this
turns out to be quite typical. The $n_w=1$ fit is very close to
the corresponding $n_w=0$ fit far from the inflection point, but the
fit becomes noticeably worse near the inflection point. This reflects
the fact that we force the fit to contain an inflection point when the
simulated foregrounds do not have one. For the rest of the paper we
therefore consider only $n_w=0$. It seems unlikely that realistic
modifications to the foreground model alone would force us to relax
this assumption. We do not know at present if, for example,
calibration errors may induce a smooth change in the change of the
spectrum which would introduce inflection points, but we know of no
specific effect which would do so.

In the remainder of this section we compare foreground subtraction
using Wp smoothing with that using parametric fitting and smoothing
splines, and study how its performance is affected by changes in the
frequency resolution and range, in the weights $c_i$ and in the model
for the CS.  We start, though, by choosing a value for the smoothing
parameter, $\lambda$, and describing the criteria we use to determine
the quality of the fitting.

\subsection{Choice of smoothing parameter}\label{subsec:chooselambda}

\begin{figure}
  \begin{center}
    \leavevmode \psfig{file=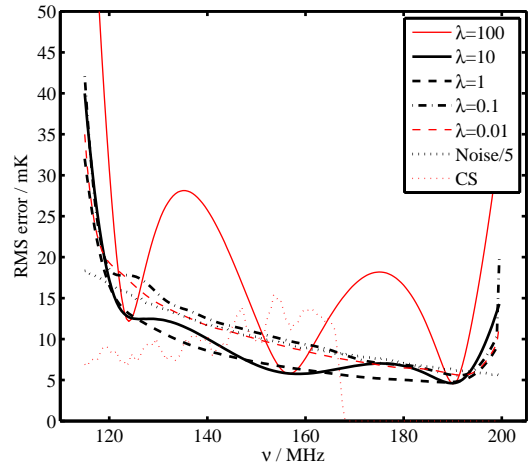,width=8cm}
    \caption{The {\it rms} difference between the known, simulated
    foregrounds, and the foregrounds estimated from Wp smoothing for
    our $256^2$ lines of sight, as a function of frequency. The solid,
    dashed and dot-dashed lines show estimates using different values
    of the smoothing parameter, $\lambda$, as given in the
    legend. With dotted lines we show the {\it rms} of the noise,
    scaled down by a factor of 5 to facilitate
    comparison, and the {\it rms} of the CS.}\label{fig:lambdarms}
  \end{center}
\end{figure}

Perhaps the most natural way to estimate the quality of the fit is to
look at the {\it rms} difference between the simulated foregrounds,
which are known exactly, and the estimates for the foregrounds
extracted from the complete data cube. We show this {\it rms}
difference as a function of observing frequency, for five different
values of $\lambda$, in Fig.~\ref{fig:lambdarms}.
We also show the frequency dependence of the noise, which we have
scaled down by a factor of 5 for ease of comparison, and the {\it rms}
of the CS. The fact that we must scale the noise for this comparison
shows immediately that the fitting errors are much smaller in
magnitude than the noise. Indeed, this is a relatively easy target to
achieve with parametric or non-parametric fits, and is achieved for
all the values of $\lambda$ shown. Good `by eye' fits are also easy to
obtain for individual lines of sight. Other than for $\lambda=100$,
over most of the frequency range the magnitude of the fitting errors
appears to scale roughly with the noise, as one might expect. At the
edges, however, the errors become much larger, growing to
approximately twice the size of the errors in nearby interior
bins. This does not seem unreasonable: for interior points the fit is
constrained from both sides, while for edge points it is constrained
only from one side. We study edge effects in more detail in
Section~\ref{subsec:weights}.

We have chosen to show the result for $\lambda=1$, but we find that
for values of $\lambda$ near $1$ we obtain very similar results. For
example, lines for $\lambda=0.5$ or $2$ would be almost
indistinguishable.  The $\lambda=1$ line therefore represents very
nearly the minimum {\it rms} error we can achieve using this method.
For $\lambda=0.1$ (light smoothing) the fit becomes noticeably worse:
on any one line of sight, random features of the noise pull the
fitting function around too easily. This just increases the {\it rms}
error by leaking noise into the fitting errors. For $\lambda=10$
(heavy smoothing) there is also a small increase in the average {\it
rms} error compared to $\lambda=1$. Oscillations in the error are also
clearly visible, however: the excessive smoothing prevents the fitting
function from accurately taking the shape of the underlying
foregrounds, introducing an additional, systematic error in parts of
the frequency range. We will examine this in more detail below when we
study the cross-correlation of foreground maps at a given frequency
with the fitting residuals. For now, we note merely that this sort of
error is potentially more pernicious than a mere increase in the
noise, since it allows spatial fluctuations in the foregrounds to leak
into the signal.

The results for $\lambda=100$ and $\lambda=0.01$, plotted using thin
lines, are intended to indicate how the fitting behaves in the limit
$\lambda\to\infty$ and $\lambda\to 0$ respectively. For $\lambda=100$
the oscillations which are also visible in the $\lambda=10$ fit become
very large, resulting in a poor fit. This is not unexpected, since the
$\lambda\to\infty$ limit for Wp smoothing for $n_w=0$ is the
best-fitting quadratic function, which we know from previous work to
be a poor model for our data compared to, for example, a cubic
function. The $\lambda\to 0$ limit is more interesting, since it
corresponds to the best-fitting function with no inflection points in
the interval. For $\lambda=0.01$ and $\lambda=0.1$ the fits are very
similar, and do not give an {\it rms} error much worse than the best
value for $\lambda$. As we mentioned in Section~\ref{subsec:bg}, this
well behaved limit is one of the attractive features of Wp
smoothing. The poorer fit for very small $\lambda$ when compared
to $\lambda=1$ suggests that such small values allow the function too
much freedom: some of the 21cm signal and noise are fitted away
(`overfitting'). The condition that $n_w=0$ imposes quite a strong
constraint on the shape of the fits, however, and ensures that even as
$\lambda\to 0$ the {\it rms} error does not become terribly large.

The contents of Fig.~\ref{fig:lambdarms} are computed by taking an
{\it{}rms} over all $256^2$ lines of sight in our data cube. The
results for $\lambda\leq 10$ do not change appreciably if we use only,
say, $32^2$ lines of sight, and do not depend on the position of the
selected sub-region. Only the $\lambda=100$ result changes: if we
choose a sub-region where the foregrounds are relatively intense, the
size of the oscillations is reduced considerably, in some cases
producing an {\it{}rms} very similar to the $\lambda=10$ result. The
oscillations come from regions where the foregrounds are less intense,
and where a quadratic function is clearly unable to match the shape of
the foregrounds as a function of frequency. This may occur
because dim regions are where the Galactic diffuse synchrotron, which
is usually the dominant foreground component, is weakest. At the
highest frequencies, emission from sources with a flatter spectrum,
for example radio galaxies, becomes more important and can even become
dominant, leading to a flat total spectrum. This flat area can only be
fitted with a quadratic function if it is near the peak or trough of
the quadratic curve, and moreover this curve's extremum must be broad.
At the lower frequencies, where Galactic synchrotron takes over again,
the spectrum becomes steeper and cannot be fitted by a quadratic near
its (broad) extremum. This leads to systematic errors in the shape of
the fit, which manifest themselves as wiggles in the plot of the {\it
rms} error as a function of frequency.

\begin{figure}
  \begin{center}
    \leavevmode \psfig{file=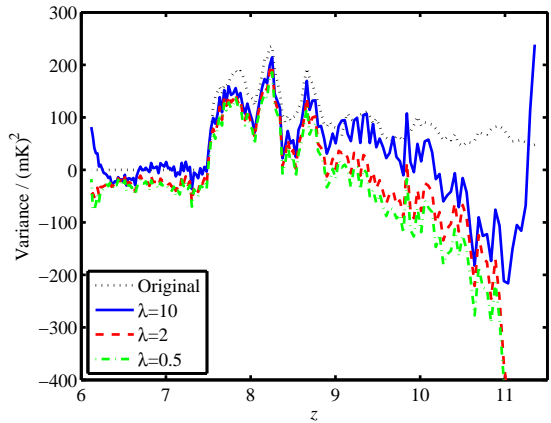,width=8cm}
    \caption{The recovered global signal from the EoR as a function of
    redshift, for three different values of $\lambda$. The variance of
    the fluctuations of the input CS is shown as the
    black, dotted line. The other three lines show estimates of this
    quantity extracted from the simulated data cube. Negative
    estimates for this variance arise because of over-fitting: the
    variance of the residuals after foreground subtraction is smaller
    in this case than the noise variance.}\label{fig:lambdaresid}
  \end{center}
\end{figure}

The first objective of the LOFAR EoR key project is simply to make a
detection of emission from the EoR, and to find the redshift evolution of
the global emission which would be a signature of reionization. If we
look at the variance of the residuals after the foregrounds have been
subtracted from the data, then subtract the (known) variance of the
noise, any remaining variance is expected to arise from fluctuations
in the CS. This change in the variance of the
fluctuations as a function of redshift constitutes a detection of the
global signature of reionization. Fig.~\ref{fig:lambdaresid} shows
how well this variance is recovered for different values of the
smoothing parameter, $\lambda$.
The black, dotted line shows the variance of the input CS, while the
other three lines show the estimates recovered from the full data
cube. We do not plot a line for $\lambda=0.1$ because it overplots the
$\lambda=0.5$ line almost exactly. For the majority of the redshift
range, $z\approx 6$--$10$, the Wp smoothing with $\lambda=10$ does
reasonably well in recovering the variance of the CS (much larger
$\lambda$, as expected from Fig.~\ref{fig:lambdarms}, does poorly, the
fitting errors adding to the {\it{}rms} of the residuals and resulting
in a large overestimate of the variance). It does better than
$\lambda=0.5$ or $2$ in this range, a property which is not clearly
reflected in Fig.~\ref{fig:lambdarms}. In this sense,
Fig.~\ref{fig:lambdaresid} does a better job of showing the effect of
over-fitting, which reduces the variance of the fitting residuals and
causes us to underestimate the CS.

\begin{figure}
  \begin{center}
    \leavevmode \psfig{file=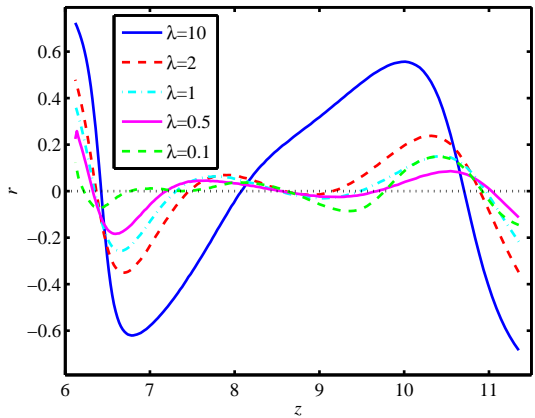,width=8cm}
    \caption{The Pearson correlation coefficient between maps of the
    input foregrounds and the corresponding maps of the fitting
    errors at the same observing frequency. We define the error to be
    the value of the fit minus the value of the input foregrounds, so
    a positive correlation implies that where the foregrounds are
    positive (negative) the fit tends to overestimate (underestimate)
    the foregrounds, while a negative correlation implies that where
    the foregrounds are positive (negative) the fit tends to
    underestimate (overestimate) the foregrounds. $r=\pm 1$
    correspond to perfect correlation and perfect anticorrelation,
    respectively. The line styles are shown in the same order in the
    legend as they appear at the far left-hand side of the
    plot.}\label{fig:lambdaxcorr}
  \end{center}
\end{figure}

By contrast, Fig.~\ref{fig:lambdaxcorr} shows the effect of
under-fitting.
Here we show the Pearson correlation coefficient, $r$, between images
of the foregrounds at a given observing frequency, and images of the
fitting errors (difference between the fit and the known foregrounds)
at the same frequency. If the pixels of the foreground image and of
the image of fitting errors have the values $a_i$ and $b_i$
respectively, where $i=1,\ldots,256^2$, then $r$ is given by
\begin{equation}
r=\frac{\sum_i(a_i-\bar{a})(b_i-\bar{b})}
{\left[\sum_i(a_i-\bar{a})^2\sum_i(b_i-\bar{b})^2\right]^\frac{1}{2}}\ ,
\end{equation}
where $\bar{a}$ and $\bar{b}$ are the mean of $a_i$ and $b_i$
respectively. $r=1$ corresponds to perfect correlation and $r=-1$ to
perfect anticorrelation. We see immediately in
Fig.~\ref{fig:lambdaxcorr} that for heavy smoothing, $\lambda=10$,
there are quite strong correlations ($r=\pm 0.6$) between the
foregrounds and the fitting errors in some parts of the frequency
range. The level of correlation reduces as $\lambda$ is reduced,
though there appears to be little to choose between $\lambda=0.5$ and
$0.1$.

This shows that, as one might expect, heavier smoothing is more likely
to allow spatial power to leak from the foregrounds into the residual
maps used for later analysis. What constitutes an acceptable level of
leakage will depend on the properties of the real foregrounds: if they
do not contain more power (especially small-scale power) than our
simulated foregrounds, and if {\it rms} errors of the order of those
shown in Fig.~\ref{fig:lambdarms} can be achieved, even correlations as
large as those shown for $\lambda=10$ in Fig.~\ref{fig:lambdaxcorr}
may not seriously harm the recovery of power spectra or other
statistics. None the less, heavy smoothing, which retains more of the
desired signal (see Fig.~\ref{fig:lambdaresid}) at the expense of
systematic correlations with the foregrounds, can be regarded as a
more aggressive foreground cleaning strategy.  Light smoothing runs
more of a risk of cleaning away the signal, but may be less
susceptible to systematics, and so may therefore be regarded as a more
conservative detection strategy.

The way the recovered variance falls away at high redshift (where the
noise is larger) in Fig.~\ref{fig:lambdaresid} seems to suggest that
more regularization is required there, i.e.\ that we may want to
consider varying $\lambda$ as a function of frequency. In practice,
doing so does not appear to deliver any significant overall
improvement in performance. We do note, though, that
equation~\eqref{eqn:minimizes} implies that a change in $\lambda$ is
degenerate with an overall scaling of the weights, and that we
directly address changes in the weighting scheme in
Section~\ref{subsec:weights}.

\begin{figure}
  \begin{center}
    \leavevmode \psfig{file=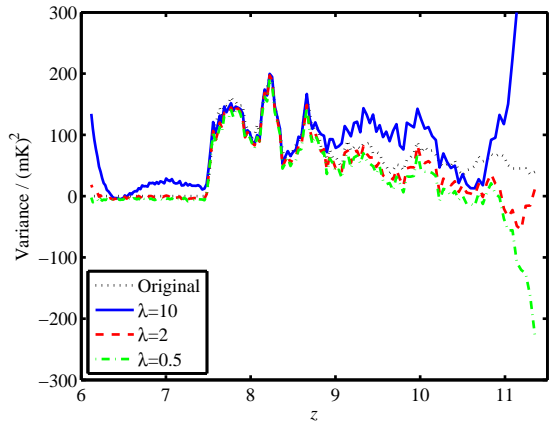,width=8cm}
    \caption{As in Fig.~\ref{fig:lambdaresid}, we show the recovered
    global signal from the EoR as a function of redshift, for three
    different values of $\lambda$. In this case, however, before
    calculating the variance of the residual maps or the maps of the
    original signal, we smooth the maps with a $4\times 4$ boxcar
    filter. The line styles are as in
    Fig.~\ref{fig:lambdaresid}.}\label{fig:lambdaresidsmooth}
  \end{center}
\end{figure}

The recovery at high redshift can be improved, however, if we look at
the variance of spatially smoothed maps. This is because the noise and
fitting errors are most dominant on small scales, and smoothing
removes small scale power. We illustrate this in
Fig.~\ref{fig:lambdaresidsmooth}.
This is very similar to Fig.~\ref{fig:lambdaresid}, apart from the
fact that before calculating the variance of the residual maps or the
maps of the original signal, we smooth them with a square boxcar
filter of $4\times 4$ pixels. The foreground fitting and subtraction
are still carried out on the full resolution data cube. For
$\lambda=0.5$ and $2$, this allows us to recover a reaonable estimate
of the variance of the CS at this scale over a much larger portion of
the redshift range than in Fig.~\ref{fig:lambdaresid}. It also
illustrates more clearly that the increase in the recovered variance
for $\lambda=10$ is spurious and comes about because of leakage of
power from the foregrounds into the residual maps. The recovered
variance exceeds the true variance of the CS at precisely those
redshifts where, as we may see from Fig.~\ref{fig:lambdaxcorr}, the
(anti-)correlations of the fitting errors with the foregrounds are
largest. Fig.~\ref{fig:lambdaresidsmooth} suggests that smoothing on
an appropriate scale may help to detect the signature of reionization
in early data from EoR experiments, and points towards the need for a
full power spectrum analysis to properly study the scale dependence of
the various components of the data cubes.  Since this paper is
concerned with the quality of the fitting, and since the variance of
unsmoothed maps seems to provide a stringent test of this, we do not
further explore scale dependence here. The recovery and analysis of
the power spectrum will instead be studied in a forthcoming paper.

\begin{figure}
  \begin{center}
    \leavevmode \psfig{file=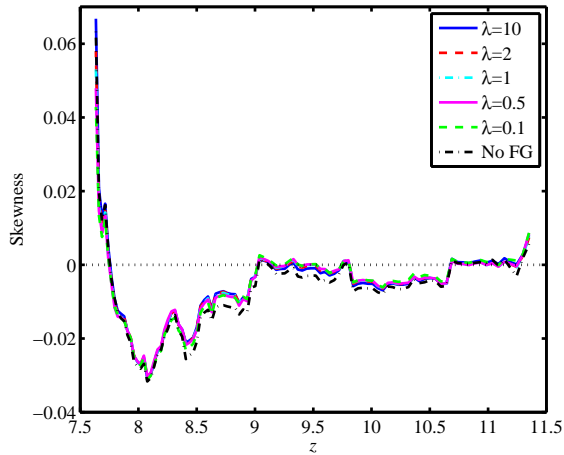,width=8cm}
    \caption{The skewness of the one-point distribution of pixel
    intensity as a function of redshift, after subtracting the
    foregrounds using Wp smoothing with different values of $\lambda$,
    and then applying a Wiener deconvolution to each image as in
    \citet{SKE_09}. We also plot the skewness we recover if we
    apply the deconvolution to maps consisting only of the signal plus
    the noise, which are equivalent to maps where the foregrounds have
    been subtracted perfectly. The lines in the figure are smoothed
    with a boxcar filter 20 points wide to improve the clarity of the
    plot.}\label{fig:wpskew}
  \end{center}
\end{figure}

In Fig.~\ref{fig:wpskew} we show how changing $\lambda$ affects our
recovery of the changes in the skewness of the one-point distribution
of the signal.
As in \citet{SKE_09} we apply a Wiener deconvolution to
foreground-subtracted images, and plot the skewness of the
distribution of pixel intensity in these images as a function of
redshift. {The lines in the plot are boxcar smoothed using a
window with a span of 20 points, so that the different histories can
be compared by eye more easily. Note that the plot does not extend to
the lowest redshift in our data cube: at the lowest redshifts the
signal is very small (c.f. the lower panel of Fig.~\ref{fig:onelos})
and the deconvolution becomes unstable. The recovered changes in
skewness are very robust to altering $\lambda$, with nearly identical
histories being produced. We also plot the result for the case of
perfect foreground subtraction (labelled `no FG' in the legend), which
is calculated by applying the deconvolution to a datacube consisting
only of the signal and noise. The foreground-subtracted cubes
reproduce this expected result quite well.

Reionization causes a fall in the skewness (to negative values in our
simulations: recall that if the neutral hydrogen density merely traced
the cosmological density field then we would expect positive
skewness), followed by a rise at low redshift \citep{WYI07,SKE_09}. We
note that the foreground simulation used here does not possess large
skewness. Thus, correlations between the foregrounds and the fitting
errors do not cause serious contamination in the recovered skewness of
the 21cm signal. If the real foregrounds turn out to be more skewed,
Fig.~\ref{fig:wpskew} suggests we would be wise to choose a value of
$\lambda$ which minimizes the correlations, perhaps at the expense of
reducing the variance of the recovered signal. Since with our current
foreground models the extracted skewness does not appear to be
sensitive to the value of $\lambda$, for the remainder of the paper we
do not use the recovered skewness to test our fitting.

There is quite a wide range of reasonable values for $\lambda$ which
achieve a compromise between over- and under-fitting. For the purposes
of comparison to other techniques in the remainder of this section we
adopt $\lambda=0.5$, since Fig.~\ref{fig:lambdaxcorr} shows there
seems to be little or no benefit from moving to smaller values (for
which the fit is slower to compute). 

To help illustrate some properties of the fitting, in
Fig.~\ref{fig:xcorrsignoise} we show the correlation coefficient, $r$,
between the fitting errors obtained using this value of $\lambda$ and
the input CS and noise.
\begin{figure}
  \begin{center}
    \leavevmode \psfig{file=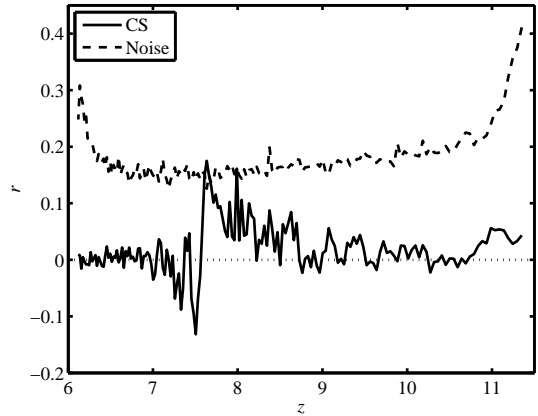,width=8cm}
    \caption{We show the Pearson correlation coefficient, $r$, between
    the fitting errors (FE) and the CS (solid line) and between the FE
    and the noise (dashed line).}\label{fig:xcorrsignoise}
  \end{center}
\end{figure}
This complements Fig.~\ref{fig:lambdaxcorr} in shedding some light on
the origin of the fitting errors. Naturally, maps of the noise are
positively correlated with maps of the fitting errors: this just shows
that where the noise is positive we tend to overestimate the
foregrounds, and vice versa. Over most of the frequency range, the
correlation coefficient between the fitting errors and the CS is close
to zero. The CS is so small compared to the noise that its effect on
the fitting errors is almost negligible. There is a feature at
$z\approx 7.5$, however, which is the redshift at which reionization
is (rather abruptly) completed in our simulation of the CS, as can be
seen in e.g.\ Fig.~\ref{fig:lambdaresid}. Despite the fact that we
cannot measure the mean 21cm signal at a given frequency using an
interferometer, the drop in the mean caused by reionization may lead
to a feature spanning several spectral bins of a particular line of
sight. Though the feature will tend to be much smaller than the noise
in any particular bin, the fact that it is correlated between bins may
mean it can affect the fitting. We caution that reionization is
completed very rapidly in this particular simulation, and that the box
size is not especially large compared to the size of individual
ionized bubbles towards the end of reionization, so that it does not
constitute a properly representative region of the Universe. Even the
small correlation we see between the signal and the fitting errors in
Fig.~\ref{fig:xcorrsignoise} may therefore be an overestimate. A
similar plot made using the other two simulations described in
Section~\ref{sec:sims}, which both produce more gradual reionization,
shows no such sharp feature.

Our final comment on Fig.~\ref{fig:xcorrsignoise} is that the fitting
errors appear to have only weak correlations with all three components
of our data cube (CS, noise and foregrounds) for Wp smoothing with
$\lambda=0.5$. At first sight this seems to make the source of the
errors somewhat obscure. However, this happens because the fitting
error at some point in some frequency bin is not caused only by the
noise at that frequency, but also by noise in nearby frequency bins,
due to the smoothing. The correlation coefficients plotted here are
calculated only between maps at the same frequency, so they do not
show this influence directly.

\subsection{Comparison to other fitting methods}\label{subsec:comparison}

We compare the performance of Wp smoothing with $\lambda=0.5$ with two
other techniques in Figs.~\ref{fig:residxcorrcomp}
and~\ref{fig:onelosfourpanel}.
\begin{figure}
  \begin{center}
    \leavevmode \psfig{file=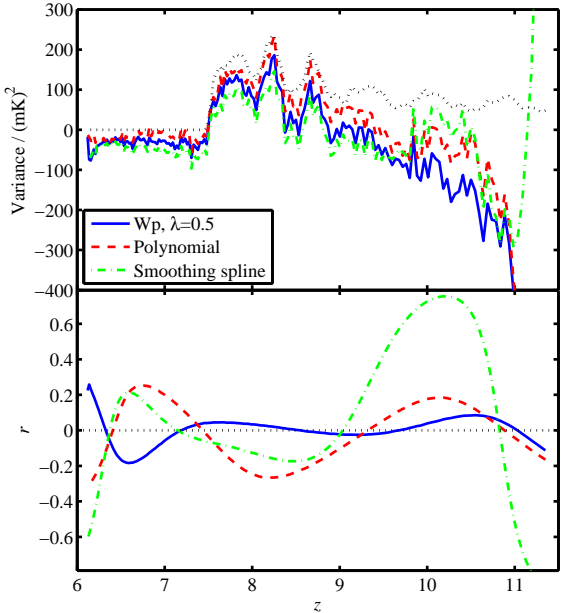,width=8cm}
    \caption{We compare the performance of Wp smoothing with
    $\lambda=0.5$ (solid blue lines) with a third-order polynomial fit
    (dashed red lines) and smoothing splines with $p=3\times 10^{-5}$
    (dot-dashed green lines; see equation~\eqref{eqn:ssdef} for a
    definition of $p$). The top panel is similar to
    Fig.~\ref{fig:lambdaresid} and shows how well each method recovers
    the variance of the fluctuations in the CS (black
    dotted line) as a function of redshift. The bottom panel is
    similar to Fig.~\ref{fig:lambdaxcorr} and shows the Pearson
    correlation coefficient between the fitting errors and the
    foregrounds.}\label{fig:residxcorrcomp}
  \end{center}
\end{figure}
The top panel of Fig.~\ref{fig:residxcorrcomp} shows how well the
three methods recover the variance of the fluctuations in the CS as a
function of redshift, as in Fig.~\ref{fig:lambdaresid}, while the
bottom panel shows the Pearson correlation coefficient between the
fitting errors and the foregrounds, as in
Fig.~\ref{fig:lambdaxcorr}. The four panels of
Fig.~\ref{fig:onelosfourpanel} show the fitting errors for four
different lines of sight. The line styles are the same for both
figures: the solid blue lines show the Wp results, the red dashed
lines show the results when we estimate the foregrounds by fitting a
third-order polynomial in $\log \nu$ to each line of sight, and the
green dot-dashed lines show the results using smoothing splines to fit
the foregrounds. Smoothing splines are a non-parametric method which
we considered as an alternative to Wp smoothing. The smoothing spline
fit is a piecewise polynomial function $f$ minimizing
\begin{equation}
p\sum_i^n
c_i[y_i-f(x_i)]^2+(1-p)\int_{x_1}^{x_n}[f''(x)]^2\mathrm{d}x\ ,
\label{eqn:ssdef}
\end{equation}
where $p$ is a smoothing parameter. $p=0$ gives a straight line fit,
while for $p=1$ $f$ becomes an interpolating cubic spline. For
Figs.~\ref{fig:residxcorrcomp} and~\ref{fig:onelosfourpanel} we used
$p=3\times 10^{-5}$.

Fig.~\ref{fig:residxcorrcomp} suggests that the smoothing spline fit
does poorly compared to the Wp smoothing: not only does it suppress
the variance of the residuals more than Wp smoothing for our chosen
values of $\lambda$ and $p$ over most of the frequency range (a
symptom of over-fitting), but it simultaneously produces fitting
errors which correlate more strongly with the foregrounds (a symptom
of under-fitting). For a small frequency interval near $z=10$, the
smoothing spline fit appears to suppress the variance less than the
other methods. This, however, is precisely the interval where the
correlations of the errors with the foregrounds are strongest, which
illustrates our point about the dangers of foreground
leakage. Similarly to the Wp case, we can improve the performance of
the smoothing spline fits according to either the over-fitting or
under-fitting criterion by tuning $p$, but this comes at the expense
of worse performance according to the other criterion. The fit is
rather sensitive to changes in $p$. For $p=3\times 10^{-6}$, a factor
of ten smaller than the value used for Fig.~\ref{fig:residxcorrcomp}
(recall that for the smoothing spline fit, smaller p corresponds to
heavier smoothing), the fitting errors become almost perfectly
correlated with the foregrounds for a large part of the interval
around $z=9.5$. This causes the recovered variance to shoot off the
scale of the top panel of Fig.~\ref{fig:residxcorrcomp}, because of
power leaked from the foregrounds. If we instead use $p=3\times
10^{-4}$, ten times larger than the value used in
Fig.~\ref{fig:residxcorrcomp}, the overfitting becomes so severe that
the recovered variance is positive for only 12 of the 170 frequency
channels. Such sensitivity to the value of the smoothing parameter may
reflect the fact that neither $p\to 0$ nor $p\to 1$ results in a
functional form that matches the expected foregrounds at all
well. Even in the best case, shown in Fig.~\ref{fig:residxcorrcomp},
the smoothing spline fit performs worse than our best Wp smoothing fit
according to both of our chosen criteria. Wp smoothing therefore
appears to be a superior method for this problem.

Comparison to the parametric (third order polynomial) fit gives a more
mixed result. For $\lambda=0.5$ the Wp smoothing loses more of the
signal, but induces smaller correlations between the fitting errors
and the foregrounds. Wp smoothing does, though, give us the freedom to
change $\lambda$ continuously to trade off performance in these two
tests. A similar trade-off is possible by changing the order of the
polynomial used for the parametric fit, but changing the order in this
way corresponds to a rather drastic jump in the properties of the fit,
and seems not to be very useful in practice. We must also emphasize
that by using Wp smoothing we are only making rather general
assumptions about the smoothness of the foregrounds (and, for our
current choice of implementation, the number of inflection points of
the foregrounds). Clearly, if we were to know the functional form of
the foregrounds in advance then we would be justified in
parametrically fitting the foregrounds with the correct function,
and we could doubtless find a parametrized form which would fit our
particular simulated foregrounds better. If, though, we can achieve
comparable results for realistic simulated foregrounds using
parametric or non-parametric methods, it would be preferable to use
the non-parametric technique on the observational data in case the
real foregrounds do not match our expectations. The fact that Wp
smoothing can achieve a fit of parametric quality without assuming a
functional form for the foregrounds justifies its use for EoR
experiments, and suggests further investigation of non-parametric
techniques to address this problem.

\begin{figure}
  \begin{center}
    \leavevmode \psfig{file=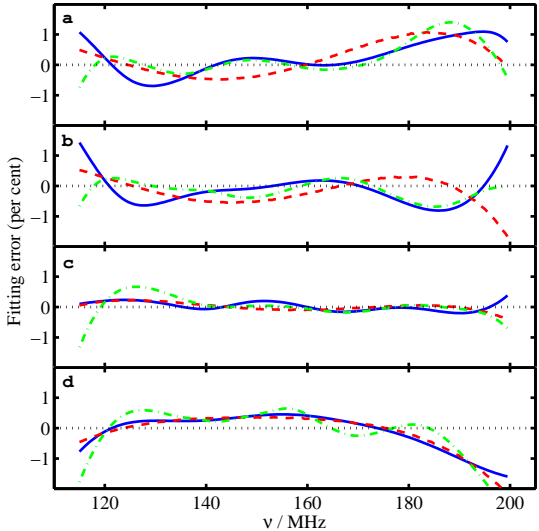,width=8cm}
    \caption{We show the fitting errors along four example lines of
    sight for Wp smoothing with $\lambda=0.5$, a third-order
    polynomial fit and smoothing splines with $p=3\times
    10^{-5}$. Line styles are as for
    Fig.~\ref{fig:residxcorrcomp}. From top to bottom, the level of
    the foregrounds at $150\ \mathrm{MHz}$ for each of the lines of
    sight is $1.89$, $1.65$, $4.93$ and $-1.14\ \mathrm{K}$. The top
    panel shows the same sight line as
    Fig.~\ref{fig:onelos}.}\label{fig:onelosfourpanel}
  \end{center}
\end{figure}
The four example lines of sight shown in
Fig.~\ref{fig:onelosfourpanel} are intended to illustrate some of the
differences between the methods. The foregrounds differ in amplitude
between these lines of sight: from top to bottom, their value at $150\
\mathrm{MHz}$ is $1.89$, $1.65$, $4.93$ and $-1.14\
\mathrm{K}$. Comparing panels a and b, one may notice that the shape
of the error curve for the polynomial fitting is very similar in these
two cases, while the Wp smoothing curve differs between the two panels
at the high frequency end. This is a manifestation of the systematic
errors made by the parametric fit which seem to be alleviated somewhat
by non-parametric methods. The line of sight in panel c comes from a
point on the sky where the foregrounds are relatively intense. The
noise does not scale with the foregrounds, and so the fitting is able
to determine the foregrounds more accurately in a relative sense. This
suggests that the large amplitude of the foregrounds relative to the
CS may be less of a concern than the scale-dependence of their
fluctuations on the sky, since small-scale fluctuations which leak
into the residual maps because of biased fitting may be confused with
the CS. Finally, panel d of Fig.~\ref{fig:onelosfourpanel} shows how
the fits produced by the smoothing spline method are more prone to
oscillations than those produced by Wp smoothing or by polynomial
fits. The statistical signature of these oscillations is the
over-fitting shown by the top panel of
Fig.~\ref{fig:residxcorrcomp}. One must be careful not to
over-interpret results for individual lines of sight, however, and so
in the remainder of the paper we restrict ourselves to statistical
comparisons.

We should note that there are, of course, many other
non-parametric methods for fitting data, or for removing noise to
reveal the smooth, underlying trends. We have only briefly
investigated some of these -- such as local regression and wavelet
denoising -- since early results suggested that the overfitting
problem is very severe when compared with Wp smoothing or with
smoothing splines, to the extent that it can be hard to compare the
results on the same figures. This is one of the problems which has made
non-parametric techniques appear unpromising for EoR foreground
subtraction until now.
 
\subsection{Changes in frequency resolution and
  weighting}\label{subsec:weights}

\begin{figure}
  \begin{center}
    \leavevmode \psfig{file=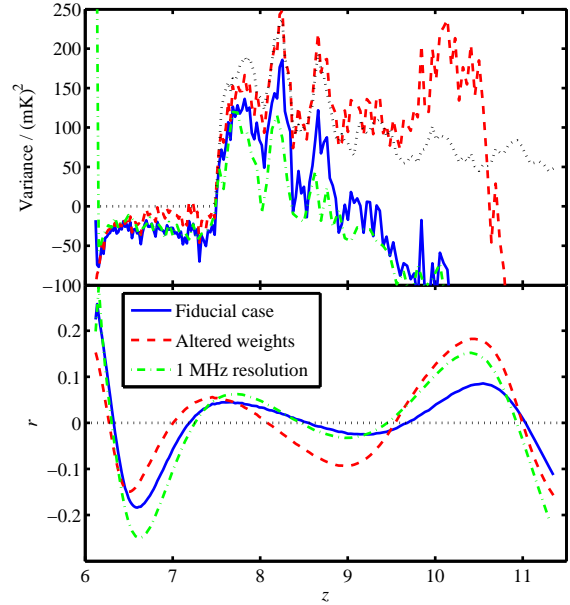,width=8cm}
    \caption{We show an example of the effects of a different
    weighting scheme and a lower frequency resolution on the recovery
    of the variance of the CS (top panel), and on the
    correlation between the fitting errors and the foregrounds (bottom
    panel). The solid blue line shows the results for $\lambda=0.5$
    and our fiducial weighting scheme and frequency resolution. The
    dashed red line shows the result if we adjust our weighting scheme
    to give points near the ends more weight, while the dot-dashed
    green line shows the effect of halving the frequency
    resolution. Note that the axes cover a smaller range than in
    Fig.~\ref{fig:residxcorrcomp}.}\label{fig:weightshrcomp}
  \end{center}
\end{figure}
It is very noticeable in Fig.~\ref{fig:lambdarms} that the errors on
the fit become larger at the ends of the frequency range. Similarly,
in Fig.~\ref{fig:lambdaxcorr}, while there is a very small
cross-correlation between the foregrounds and fitting errors for
$z\approx 7$--$10$ for our $\lambda=0.5$ fit, the performance degrades
slightly at the lowest redshifts (highest frequencies). It would be
desirable to have a fit of more uniform quality, since otherwise we
truncate the useful frequency range, and since we might worry that an
apparent signal is merely a side-effect of more serious foreground
contamination at some redshifts than others. It seems plausible that
adjusting the weights $c_i$ used in the fitting may improve the fit at
the ends of the interval at the expense of the interior. Modest
changes in the weights (for example, using uniform weights rather than
inverse noise weights) have little effect. Large enough changes do
have an impact, though, as we show in Fig.~\ref{fig:weightshrcomp}.
Here we compare our fiducial weighting scheme (solid blue line) with
an alternative weighting scheme (dashed red line) in which extra
weight is given to points near the ends of the interval. To be
precise, we multiply the $i$th `natural' weight $1/\sigma_i$ by
$1/(1-d_i^2)$ where $d_i=1.7(i-1)/(n-1)-0.9$. We then normalize the
new weights to have the same mean as the fiducial weights, in order
that the value of $\lambda$ can remain unchanged. The top panel of the
figure shows the recovered variance, while the bottom panel shows the
correlation coefficient between fitting errors and foregrounds, as in
Fig.~\ref{fig:residxcorrcomp}.

It seems that this adjustment of the weighting scheme is at least a
limited success. The correlation between fitting errors and
foregrounds becomes slightly smaller at low redshift, at the expense
of increased correlations in the interior of the redshift range. The
recovered variance of the signal is, moreover, closer to the original
in the most interesting part of the redshift range. Unfortunately, the
origin of this improved agreement is not a better fit, but a worse
one. This is demonstrated in Fig.~\ref{fig:rmsplotweights}, in which we
show the {\it rms} error of the foreground fitting. The line styles
are the same as for Fig.~\ref{fig:weightshrcomp}.
\begin{figure}
  \begin{center}
    \leavevmode \psfig{file=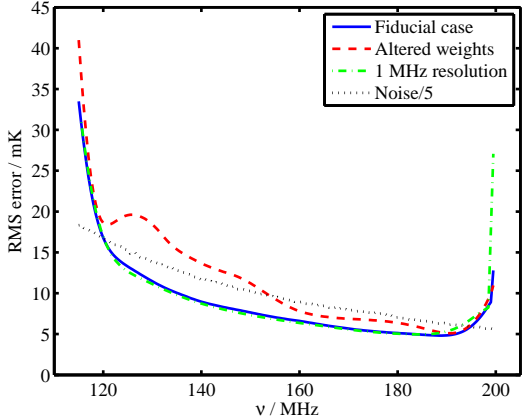,width=8cm}
    \caption{We show the {\it rms} error of the foreground fitting for
    our fiducial fit, a fit with modified weights, and a fit for a
    data cube with half the frequency resolution of our standard data
    cube. The line styles are as for Fig.~\ref{fig:weightshrcomp}. The
    solid and dot-dashed lines almost coincide for most of the
    frequency range.}\label{fig:rmsplotweights}
  \end{center}
\end{figure}
The modified weighting scheme significantly increases the fitting
errors. The improved recovery of the signal variance in
Fig.~\ref{fig:weightshrcomp} therefore seems to be a fluke caused by
leaking more noise into the fitting residuals, and it would be hard to
recommend this as a strategy for signal recovery. In fact, after
experimenting with various weighting schemes, modifying them seems to
be an unpromising avenue: modest changes have a marginal effect, while
large changes tend to significantly increase the overall error.

Also shown in Fig.~\ref{fig:weightshrcomp} is the effect of reducing
the frequency resolution to $1\ \mathrm{MHz}$ rather than $0.5\
\mathrm{MHz}$. Though this halves the number of bins, it also reduces
the noise per bin by a factor of $\sqrt{2}$. In the top panel we see
that the recovered variance is smaller, but this is not due to a
poorer fit being achieved (as one can see from
Fig.~\ref{fig:rmsplotweights}): rather, the variance of the original
signal itself is reduced when binned up, since adjacent $0.5\
\mathrm{MHz}$ frequency slices are decorrelated to some extent. The
amount of variance lost by the fitting process is similar in either
case. The reduction in the number of data points does, however,
degrade the quality of the fit in the sense that the correlation
between fitting errors and foregrounds increases, as one can see in
the lower panel of Fig.~\ref{fig:weightshrcomp}. Increasing the number
of frequency channels stored and analysed may be expensive,
unfortunately. Since we can achieve low foreground contamination in
our $0.5\ \mathrm{MHz}$ case, a further increase in frequency
resolution may only significantly reduce the fitting contamination if
a smaller frequency range is being observed and so a larger number of
bins is required to avoid edge effects. Otherwise, a more stringent
criterion for selecting the frequency resolution would be to choose it
such that the decorrelation within a resolution element is not too
large.

\subsection{Alternative signal models and frequency
  ranges}\label{subsec:altsims}

So far we have shown results using only the f250C simulation of
\citet{ILI08}. We now show the effect on the signal extraction of
taking our CS from the two simulations, T-QSO and
T-star (see Section~\ref{sec:sims}) described by \citet{THO09}. The
top panel of Fig.~\ref{fig:orxs} shows the variance of the
CS derived from each of these three simulations as a
function of redshift.
\begin{figure}
  \begin{center}
    \leavevmode \psfig{file=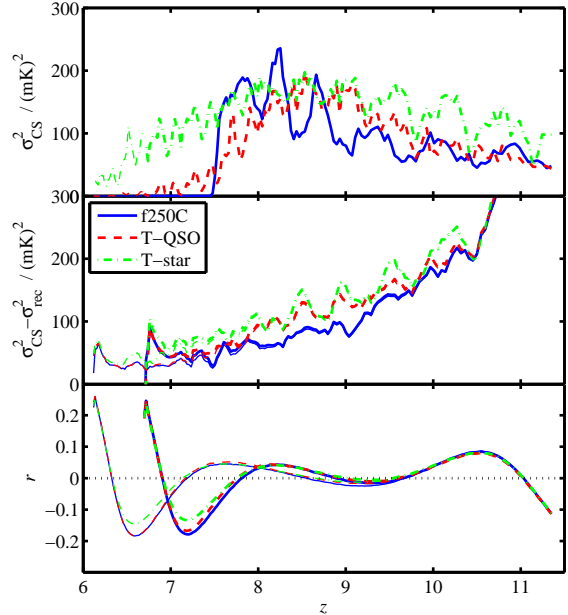,width=8cm}
    \caption{We study the effect on our extraction of using a
    different model for the CS and of truncating the
    frequency range used for the fit. Thin lines show results using
    our normal frequency range, $\nu=115$--$200\ \mathrm{MHz}$, while
    thick lines show results using $\nu=115$--$185\ \mathrm{MHz}$. In
    both cases the frequency resolution is $0.5\ \mathrm{MHz}$. The
    top panel shows the variance in three different simulations of the
    CS as a function of redshift. The solid blue line
    uses the f250C simulation of \citet{ILI08} which we have been
    using throughout the paper, and is the same as the dotted line of
    Fig.~\ref{fig:lambdaresid}. The red dashed line and the green
    dot-dashed line are for the simulations of \citet{THO09} which
    assume that reionization is carried out by QSOs and by stars,
    respectively. The middle panel uses the same colour coding, and
    shows the difference between the recovered variance and the true
    variance of the CS. The bottom panel shows the
    Pearson correlation coefficient between the fitting errors and the
    foregrounds.}\label{fig:orxs}
  \end{center}
\end{figure}
This variance goes to zero at low redshift as reionization destroys
the neutral hydrogen responsible for 21cm emission. The speed of this
decline varies between simulations. The solid blue line (f250C) is
most rapid, followed by the dashed red line (T-QSO) then the
dot-dashed green line (T-star). This set of simulations is therefore
useful to check that the quality of our fits is not unduly influenced
by details of the signal. We also use all three simulations to test
our procedure over a shorter frequency interval, from $115$ to $185\
\mathrm{MHz}$ ($z=11.35$--$6.70$) rather than our fiducial
$115$--$200\ \mathrm{MHz}$ ($z=11.35$--$6.12$). We aim to check
whether the different low redshift (high observing frequency)
behaviour leads to this truncation having a different effect. This is
an important test because it may not be possible to observe over the
entire frequency range at once with LOFAR. Rather, we may have to
split the frequency range into $32\ \mathrm{MHz}$ chunks which are
observed consecutively. It may then be necessary to choose between
increasing observing time at, say, $115$--$180\ \mathrm{MHz}$ and
increasing the frequency range to $115$--$210\ \mathrm{MHz}$.

The middle and bottom panels of Fig.~\ref{fig:orxs} test the quality
of the fit. The colour coding is the same as for the top panel. The
thick lines show the results for an analysis using only $115$--$185\
\mathrm{MHz}$ while the thin solid lines show our fiducial
$115$--$200\ \mathrm{MHz}$ case. In the middle panel we show the
difference between the recovered variance of the CS and the original
variance from the datacube without noise or foregrounds. For the thin
solid line (f250C, $115$--$200\ \mathrm{MHz}$), this is equal to the
difference between the dotted line and the solid blue line in the top
panel of Fig.~\ref{fig:weightshrcomp}; that is, it shows the amount of
variance lost through overfitting. We see that the three different
simulations show similar behaviour, though our procedure performs
slightly better for f250C than for the other two simulations. For the
majority of the frequency range, the thick and thin lines are
indistinguishable, meaning that the effects of truncating the
frequency range seem to be limited to the edge regions in this
case. In the bottom panel we plot the Pearson correlation coefficient
between the fitting errors and the foregrounds. The results using the
three simulations are again very similar. The effect of the truncation
is visible for a larger part of the range than was the case in the
middle panel, but the correlations do not become larger: rather, the
whole pattern just appears to be squashed.

We find that for larger values of $\lambda$ (heavier smoothing) the
effect of truncation on the correlation extends over a larger part of
the frequency range. Recall that larger values of $\lambda$ give a
more aggressive signal recovery strategy, and one which allows us to
detect an excess from the CS to higher redshift (that is, the lines of
Fig.~\ref{fig:lambdaresid} do not fall away as rapidly at high
redshift if $\lambda$ is large). If we wish to pursue such an
aggressive strategy, or indeed if it turns out to be necessary to do
so to detect the signal, then extending to higher frequencies may turn
out to be required: heavier smoothing makes more use of the longer
lever arm provided by extending the range of the fit.

For our fiducial value of $\lambda$, though, we infer from
Fig.~\ref{fig:orxs} that if Wp smoothing is used to fit the
foregrounds, shortening the frequency interval should not affect the
quality of signal recovery in the interior too badly, either for
extended or rapid reionization. The most important consideration when
choosing what range of frequencies to observe is that we should be
prepared to discard (or view with considerable caution) some bins at
either end of the frequency range after fitting, since they are likely
to be corrupted by edge effects. We should like to avoid discarding
bins which are likely to have an interesting contribution from the CS:
moving from an upper frequency limit of $180\ \mathrm{MHz}$ to one of
$210\ \mathrm{MHz}$ could well be advantageous in this respect, though
to some extent this will depend on the properties of the signal we aim
to find.

\section{Conclusions}\label{sec:conc}

We have argued that without a good reason to assume that the
foregrounds for EoR experiments have a specific functional form, it is
preferable to fit them with non-parametric methods that use their
assumed smoothness directly, rather than to fit parameters of some
chosen model. Unfortunately, most non-parametric methods tend to give
poor quality fits compared to parametric ones that use the `correct'
model. We suggest that Wp smoothing may be an exception to this rule
in the case examined in this paper.

Wp smoothing penalizes changes in curvature. In the general case it
does so primarily by penalizing the existence of inflection points,
but in the case that the inflection points are known or fixed, it
penalizes the integrated change of curvature `apart from inflection
points'. We have drawn attention to the results of
\citet{MAC93,MAC95}, who derives a boundary value problem the solution
of which is the desired smoothing function. We have sketched two
algorithms which suffice to solve this problem in the case of EoR
foregrounds, which we assume have no inflection points (as would be
the case for a sum of several power law spectra with negative
index). Our preferred algorithm is detailed in
Section~\ref{subsec:alg}, while the other is outlined in
Appendix~\ref{sec:appa}.

We have tested Wp smoothing on synthetic data cubes which include
contributions from a detailed simulation of the CS from the EoR, a
realistic model of the diffuse foregrounds, and the levels of noise
and instrumental corruption expected for the LOFAR EoR
experiment. Though Wp smoothing is considered to be non-parametric, it
does require the specification of a smoothing parameter which governs
the relative importance of the sum of the squared residuals and the
curvature penalty function in the fitting. For the purposes of most of
our tests we have adopted a value for $\lambda$ which, for our
dataset, provides a good compromise between over-fitting, which causes
an underestimate of the variance of the CS, and under-fitting, which
causes positive or negative correlations between the fitting errors
and the foregrounds. Using this value of $0.5$ for $\lambda$, we found
that Wp smoothing easily outperforms other non-parametric methods we
have tried, including the smoothing splines shown in
Section~\ref{subsec:comparison}, and is competitive with parametric
fitting even when we are able to choose a parametrized functional form
with advance knowledge of the foregrounds.

No scheme seems able to prevent the quality of the fit from degrading
at the ends of the frequency interval used for observation. This
problem can be mitigated somewhat by analysing data cubes with a high
frequency resolution, though we note that high resolution is already
desirable to avoid averaging away our signal, and this may be a more
important criterion when deciding what resolution to use. We can make
the quality of the fit marginally more uniform by increasing the
weight given to data points near the ends of the frequency range. We
argue, though, that the cost of doing so (in terms of increasing the
noise on the fit) is too heavy for it to be worthwhile.

It may therefore be helpful to extend the range of frequencies
observed. It is difficult to extend to lower frequencies (higher
redshifts) because of the presence of the FM band. The increasing
foreground and noise amplitude may also limit the usefulness of low
frequency observations, though it is plausible that observations with
the LOFAR low band antennas (which can observe at $30$--$80\
\mathrm{MHz}$) could help constrain the shape of the
foregrounds. Extending to higher frequencies is more
promising. Firstly, the foregrounds and noise are smaller in
amplitude. Secondly, because higher frequencies correspond to $z<6$ we
expect a negligible contribution from redshifted 21cm emission
there. This helps to establish a baseline against which we can detect
a higher redshift excess coming from the CS, and ensures that this
excess occurs well away from the problematic edges of the frequency
range. We have tested the quality of our fitting using two alternative
simulations of the CS which exhibit more extended reionization, and
have analysed all three simulations using a datacube which extends
only to $185\ \mathrm{MHz}$ rather than $200\ \mathrm{MHz}$. We find
that away from the edges, neither change badly affects the quality of
the foreground fitting.

We also note that we have concentrated primarily on the recovery of
the excess variance coming from the CS as a measure of the quality of
our fits. Other statistics such as the skewness may be more robust
\citep{SKE_09}. It is also the case that the power from fitting
errors, noise and the CS peaks at different scales, so a power
spectrum analysis may improve prospects for detection of a signal, as
well as giving more sensitive constraints on models than the
integrated variance once a detection is made (e.g. \citealt{MOR06};
\citealt*{BOW07}). Given this scale dependence, it is interesting to
consider whether or not it may be advantageous to fit out the
foregrounds in the {\it uv} plane. This has been considered
recently by \citet{LIU09b} in the context of linear least-squares
fitting and was found to afford substantial benefits, particularly at
small angular scales where foreground fluctuations arising from
unsubtracted point sources interact with the frequency-dependent
`frizz' on the outskirts of the point spread function
\citep*{LIU09a}. It is not yet clear how this method generalizes to
nonlinear fitting: the complications include, for example, that we
must fit a complex function of frequency at each point in the {\it uv}
plane, as opposed to a real function at each point in the image
plane. It is possible, though, that by adapting the fitting according
to the relative strength of the foregrounds, noise and signal at
different scales we can improve sensitivity. We defer detailed study
of power spectrum estimation and {\it uv} plane effects to future
work.

Our results suggest that by paying close attention to the method used
in fitting the foregrounds for EoR experiments, the sensitivity of
these experiments can be increased, and we may have greater confidence
that a detection of the signal is not affected too severely by
foreground contamination. Foreground subtraction is very unlikely to
be a bottleneck in the data processing and analysis pipeline, and so
it is reasonable to consider relatively sophisticated and
computationally expensive fitting methods if they provide a
benefit. We have argued that Wp smoothing does seem to provide such a
benefit, and will continue to test its performance as more elaborate
models of the foregrounds and the instrument become available.

\section*{Acknowledgments}

GH is supported by a grant from the Netherlands Organisation for
Scientific Research (NWO). As LOFAR members, the authors are partially
funded by the European Union, European Regional Development Fund, and
by `Samenwerkingsverband Noord-Nederland', EZ/KOMPAS. The authors
would also like to thank the anonymous referee for some helpful
comments, including a suggestion as to the reason for the poor fits
obtained using a quadratic function in areas of weak emission.

\bibliography{allbib}

\appendix 

\section{Alternative solution methods}\label{sec:appa}

It is possible to rewrite equations~\eqref{eqn:f2deriv}, \eqref{eqn:mainode},
\eqref{eqn:bcdef1} and \eqref{eqn:bcdef2} in a convenient form to solve
them using a standard BVP solver. We have implemented Wp smoothing in
this manner to test our finite difference scheme, and present the
equations in appropriate form here for completeness.

At first sight the boundary conditions of equation~\eqref{eqn:bcdef2} look
awkward, since they use the value of the function at points which are
not at the ends of the interval.  Solvers for such `multi-boundary'
problems are available, however.  Moreover, by reexpressing the sums
as integrals, we can take care of the boundary conditions by adding
two more differential equations to the system, in line with the
elegant trick suggested in section~5 of \citet{ASC81}.

This is promising, but doesn't help with the dependence on $f(x_i)$
for all $i$ on the right-hand side of equation~\eqref{eqn:mainode}, so
we use a different trick. We start by rewriting
equations~\eqref{eqn:f2deriv} and~\eqref{eqn:mainode} as coupled first-order
equations, as is commonly done:
\begin{align}
h'(x)&=g(x)\ , \label{eqn:syshd} \\ 
g'(x)&=p_\mathbf{w}(x)\mathrm{e}^{h(x)}\left[-\frac{1}{2\lambda}\sum_{i=1}^{n}(x-x_i)_{\scriptscriptstyle +}\psi_i(y_i-f(x_i))\right]\
, \label{eqn:sysgd} \\
f'(x)&=k(x) \ ,\label{eqn:sysfd} \\
k'(x)&=p_\mathbf{w}(x)\mathrm{e}^{h(x)} \ .\label{eqn:syskd}
\end{align}
Equations~\eqref{eqn:syshd} and~\eqref{eqn:sysfd} define our new functions
$g$ and $k$ respectively, and the boundary condition of
\eqref{eqn:bcdef1} becomes $g(x_1)=g(x_n)=0$.

Now, again following \citet{ASC81}, we split the domain of solution
into $n-1$ intervals, $[x_1,x_2]$, $[x_2,x_3]$, \ldots,
$[x_{n-1},x_n]$. In each interval we change variables, letting
\begin{equation}
t=\frac{x-x_m}{x_{m+1}-x_m}\quad \mathrm{for}\quad x_m\leq x\leq x_{m+1}
\end{equation}
which maps each interval onto the unit interval, $[0,1]$. Then, on
this interval, we define functions $f_m(t)$, $g_m(t)$, $h_m(t)$,
$k_m(t)$, $p_{\mathbf{w},m}(t)$ for $m=1,2,\ldots,n-1$ such that, for
$x_m\leq x\leq x_{m+1}$, $f_m(t)=f(x)$, $g_m(t)=g(x)$, $h_m(t)=h(x)$,
$k_m(t)=k(x)$ and $p_{\mathbf{w},m}(t)=p_\mathbf{w}(x)$. We further
define the functions $q_m(t)$ for $m=1,\ldots,n$ where $q_m(t)=f_m(0)$
for $m=1,\ldots,n-1$ and $q_n(t)=f_{n-1}(1)$. Our system of four
equations~\eqref{eqn:syshd}--\eqref{eqn:syskd} then becomes the following
system of $5n-4$ equations (where dashes now indicate differentiation
with respect to $t$):
\begin{align}
f_m'(t)={}&(x_{m+1}-x_m)k_m(t)\ , \label{eqn:tsysf} \\
k_m'(t)={}&(x_{m+1}-x_m)p_{\mathbf{w},m}(t)\mathrm{e}^{h_m(t)}\ , \label{eqn:tsysk}
\\
h_m'(t)={}&(x_{m+1}-x_m)g_m(t)\ , \label{eqn:tsysh} \\
g_m'(t)={}&(x_{m+1}-x_m)p_{\mathbf{w},m}(t)\mathrm{e}^{h_m(t)}\nonumber \\
&\times\left\{\frac{-1}{2\lambda}\sum_{i=1}^m[x_m+(x_{m+1}-x_m)t]\psi_i(y_i-q_i(t))\right\}\ ,
\label{eqn:tsysg} \\
q_j'(t)={}&0\ , \label{eqn:tsysq}
\end{align}
where the index $m$ runs from 1 to $n-1$ and $j$ runs from 1 to
$n$. The functions $q_j$ carry the value of $f$ at the data points,
$f(x_i)$, to the interior of the intervals, a property which is
imposed with the boundary conditions
\begin{align}
q_m(0)&=f_m(0)\quad \mathrm{for}\quad m=1,\ldots,n-1 ;
\label{eqn:tbcqm} \\
q_n(0)&=f_{n-1}(1) \label{eqn:tbcqn}\ .
\end{align}
Our original boundary conditions become
\begin{gather}
g_1(0)=g_{n-1}(1)=0 \ ;\label{eqn:tbcg} \\ \sum_{i=1}^n \psi_i(y_i -
q_i(0))=\sum_{i=1}^n x_i \psi_i(y_i - q_i(0))=0 \ . \label{eqn:tbcsum}
\end{gather}
The remaining $4(n-2)$ boundary conditions come from imposing
continuity on the functions $f(x)$, $g(x)$, $h(x)$ and $k(x)$:
\begin{align}
f_m(1)&=f_{m+1}(0)\ , \label{eqn:tbcfcont} \\
g_m(1)&=g_{m+1}(0)\ , \label{eqn:tbcgcont} \\
h_m(1)&=h_{m+1}(0)\ , \label{eqn:tbchcont} \\
k_m(1)&=k_{m+1}(0)\ , \label{eqn:tbckcont}
\end{align}
where here the index $m$ runs from 1 to $n-2$.

Note that the boundary conditions only involve the value of functions
at $t=0$ and $t=1$, and that to calculate the derivatives given by
equations~\eqref{eqn:tsysf}--\eqref{eqn:tsysq} at a given value of $t$ only
requires the evaluation of functions at the same value of $t$. The
system is therefore suitable for solution using the \textsc{matlab}
routine `bvp4c' \citep{KIE01}, a BVP solver that uses a collocation
method.  We call it with an initial mesh of five evenly spaced points,
and with initial conditions calculated in a similar fashion to those
used for the finite difference scheme in the main text. The system of
equations is greatly expanded from the four with which we started
since the special form of the problem is not exploited, and typically
takes several seconds to solve on our test machines.

\end{document}